\documentclass[preprint]{aastex}

\newcommand{\Msun}{\mbox{M$_{\odot}$}}

\def\HII        {\hbox{H \small{II}}}

\usepackage{graphicx}
\usepackage{txfonts}
\usepackage{natbib}

\begin{document}
\title{Interferometric mapping of magnetic fields: The ALMA view of the massive star forming clump W43-MM1}

\author{Paulo C.\,Cortes\altaffilmark{1,2}, Josep M. Girart\altaffilmark{3,4}, Charles L. H. Hull\altaffilmark{4,11}, 
Tirupati K. Sridharan\altaffilmark{4}, Fabien Louvet\altaffilmark{5}, Richard Plambeck\altaffilmark{6},
Zhi-Yun Li\altaffilmark{7}, Richard M. Crutcher\altaffilmark{8}, \& Shih-Ping Lai\altaffilmark{9,10}}

\affil{National Radio Astronomy Observatory, Charlottesville, VA 22903, USA}
\affil{Joint ALMA Office, Alonso de Cordova 3107, Vitacura, Santiago, Chile}
\affil{Institut de Ci\`encies de l'Espai, (CSIC-IEEC), Campus UAB, Carrer de Can Magrans S/N, 08193 Cerdanyola del Vall\'es, Catalonia, Spain}
\affil{Harvard-Smithsonian Center for Astrophysics, 60 Garden St., Cambridge, MA 02138, USA}
\affil{Departamento de Astronomia - Universidad de Chile}
\affil{Astronomy Department \& Radio Astronomy Laboratory, University of California, Berkeley, CA 94720-3411, USA}
\affil{Astronomy Department, University of Virginia, Charlottesville, VA 22904, USA}
\affil{Astronomy Department, University of Illinois at Urbana-Champaign, IL 61801, USA}
\affil{Institute of Astronomy and Department of Physics, National Tsing Hua University, Hsinchu 30013, Taiwan}
\affil{Institute of Astronomy and Astrophysics, Academia Sinica, P.O. Box 23-141, Taipei 10617, Taiwan}
\affil{Jansky Fellow of the National Radio Astronomy Observatory}
\email{pcortes@alma.cl}

\begin{abstract}
\noindent Here we present the first results from ALMA observations of 1 mm polarized dust emission towards the W43-MM1 high mass
star forming clump. 
We have detected a highly fragmented filament with source masses
ranging from 14 \Msun\ to 312 \Msun, where the largest fragment, source A, is believed to be one of the 
most massive in our Galaxy. 
We found a smooth, ordered, and detailed polarization pattern throughout the filament which we used to derived 
magnetic field morphologies and strengths for 12 out of the 15 fragments detected ranging from 0.2 to 9 mG. 
The dynamical equilibrium of each fragment was evaluated
finding that all the fragments are in a super-critical state which is consistent with previously detected infalling
motions towards W43-MM1. Moreover, there are indications suggesting that the field is being dragged
by gravity as the whole filament is collapsing.
\end{abstract}


\maketitle

\section{INTRODUCTION}\label{se:INTRO}

\noindent W43-MM1 is a large and young high mass star forming clump located within the W43 region and at
5.5 kpc from the sun \citep{Motte2003,Zhang2014a}. The clump is
near $l=31^{\circ}, b=0^{\circ}$ and at an interface with an extended \HII\ region powered by a cluster
of O type and Wolf-Rayet stars.
\citep{Cesaroni1988,Liszt1995,Mooney1995}.
W43-MM1 has been well studied in continuum from 1.3 mm to 70 $\mu$m \citep{Motte2003,Bally2010}.
These studies identified a large sample of clumps in W43, from which W43-MM1 is the most massive
with an estimated mass of 2128 \Msun\ and a deconvolved size of 0.09 pc \citep{Louvet2014}.
Infalling motions have been detected towards W43-MM1 \citep{Cortes2010}
suggesting that the clump is undergoing gravitational collapse.
\citet{Cortes2006a}
made interferometric observations of polarized dust emission with BIMA,
finding an ordered pattern for the magnetic field and estimating an on-the-plane of the sky
field strength of about 1 mG. 
Recent SMA results from polarized dust emission and
CH$_{3}$CN line emission at higher angular resolution ($\sim 0.1$ pc scales) updated the
magnetic field estimate to 6 mG also computing a mass to magnetic flux ratio of about the critical value 
\citep{Sridharan2014}. Additionally from the CH$_{3}$CN emission, evidence for an embedded hot core of $\sim 300$ K
was found in W43-MM1 main clump. 
In this Letter we report the first ALMA observations of polarized dust emission towards W43-MM1. Here,
section \ref{se:OBS} report the observations, section \ref{se:CE} the continuum emission and
source extraction, section \ref{se:MF} the magnetic field properties, and
section \ref{se:sum} is the summary and discussion.

\section{OBSERVATIONS}\label{se:OBS}

\noindent ALMA observations at 1 mm (band 6) were done on May 30th, 2015 over 
W43-MM1 using $\alpha=$18:47:47.0 and $\delta=$-01:54:28.0 as phase center.
An array of 35 antennas was used reaching an angular resolution of 0.5$^{\prime \prime}$ 
($\sim$ 0.01 pc scales).  The spectral configuration was set to single spectral
windows per baseband in continuum mode with 64 channels giving 31.250 MHz as the 
spectral resolution in full polarization mode
 ($XX, YY, YX$, and $XY$). Each spectral window was centered at the standard ALMA
band 6 polarization frequencies (224.884, 226.884, 238.915, and 240.915 GHz), where 
two successful executions were done as part of the session scheme \citep{ALMATechnical}.
Calibration and imaging was done using the Common Astronomical Software Applications (CASA) 
version 4.5.

\subsection{Calibration and imaging in full polarization mode}\label{se:calib}
\noindent The ALMA antennas are equipped with receivers sensitive to linear polarization.
After the incoming radiation has gone through the feed-horn, the wave is
divided into two orthogonal components ($X$ and $Y$) by a wave splitting device \citep{ALMATechnical}.
This operation is not perfect and there is always a residual, or projection, from
one polarization onto the other which is known as the instrumental polarization, or D-terms \citep{Sault1996}.
Given that an antenna uses azimuth and elevation coordinates, the frame of the sky rotates
with respect to the antenna introducing an angular dependence which is parameterized by the parallactic angle. 
Additionally to the D-terms, the $X$ and $Y$ polarizations have different signal paths which 
introduces a relative delay between both polarizations.
Also, the interferometric calibration scheme for amplitude and phase requires the
use of a reference. This reference breaks the degeneracy intrinsic to the array and thus; 
we do not measure absolute phase values but relative ones with respect to the reference (where
phases are set to zero in both polarizations). By doing this, we introduce an additional
phase bandpass between the {\em XY} and {\em YX} cross correlations.
To calibrate all these quantities, an ALMA polarization observation samples a strong, un-resolved, 
polarized source over a certain range of parallactic angle. The polarization calibrator is sampled for 5 minutes every 
35 minutes or so (the precise time cadence is calculated by the online software at run time).
For our observations, we obtained about 100$^{\circ}$ of parallactic angle coverage for J1924-292
which was selected as polarization calibrator. Using this source, we derived solutions for 
the cross polarization delay, the {\em XY}-phase, and the D-terms\footnote{More details about the 
calibration procedure can be found in the 3c286 ALMA science verification data casa-guide at
{\small https://casaguides.nrao.edu/index.php/3C286\_Band6Pol\_Calibration\_for\_CASA\_4.3}}. 
These solutions were applied to
W43-MM1 data also using 
J1751+0939 to calibrate the bandpass, J1851+0035 to calibrate the phase,
and Titan to calibrate the flux. 
After applying the calibration tables, we imaged the data using the $clean$ CASA task
with the {\em Briggs} weighting scheme, robust number 0.5, for sidelobe robustness and the Clark deconvolution algorithm
to produce the Stokes images. The final images were produced after three self-calibration iterations
using a final solution interval of 90 seconds\footnote{In principle the solution interval when self-calibrating, is calculated based on the the sensitivity. However, we noticed that using smaller solution intervals increased the 
image noise and thus, we stopped the iteration at 90s}.

\section{CONTINUUM EMISSION}\label{se:CE}

\begin{figure}
\centering
\includegraphics[width=0.9\hsize]{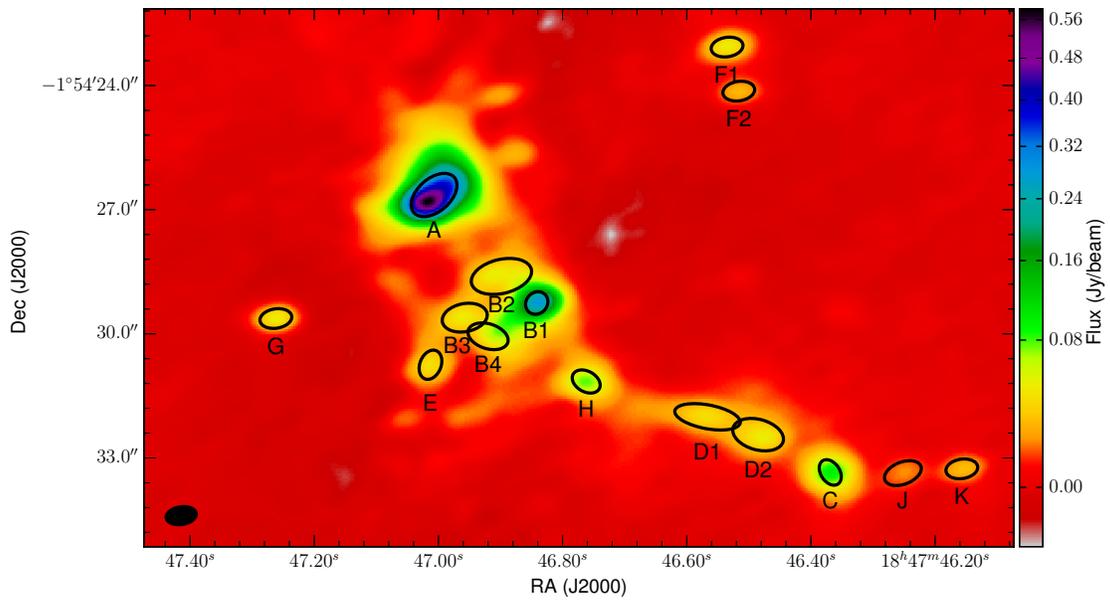}
\caption{
    The Figure shows the Stokes I emission from the W43\_MM1 clump indicated by the color-scale in Jy $\mathrm{beam}^{-1}$.
    Overlaid are the sources extracted as ellipses, in black, representing the deconvolved sized obtained from the
    Gaussian fits. With an rms noise of $\sigma = 0.41 $ mJy $\mathrm{beam}^{-1}$, the dynamic range of the
    self-calibrated image is 1220. 
}
\label{StokesI}
\end{figure}

\noindent Figure \ref{StokesI} presents the Stokes I image from W43-MM1. 
The continuum emission shows a fragmented filament extending from the north to the south-west.
Two bright sources at the center, A and B1, 
completely dominate the energy budget in W43-MM1 (with integrated fluxes of $\sim$ 2.0 and 0.5 Jy,
which correspond to $\sim 63\%$ of the total flux recovered),  
over a number of additional fragments extending to the south-west.
Also, additional sources to the east and west have been detected.
Comparing with the SMA results from \citet{Sridharan2014} and the 
PdBI 1 mm results from \citet{Louvet2014}, the ALMA 
observations reproduce quite well the overall morphology of W43-MM1, but with better resolution.
The noise in the ALMA map is $\sigma = 0.41$ mJy $\mathrm{beam}^{-1}$ with a peak
of 503 mJy $\mathrm{beam}^{-1}$ obtained from Gaussian fitting.
We used the {\em getsources}
algorithm \citep{Menshchikov2012} to  successfully extract 14 sources from
our ALMA data (see Table \ref{tab1} and Figure \ref{StokesI}). The extraction was later compared to 
other methods such as {\em clumpfind} \citep{Williams1994}, {\em FellWalker}, and {\em Reinhold} \citep{Berry2007} 
obtaining a good agreement with the
selection produced by {\em getsources}. We used the source positions and sizes derived using {\em getsources} as
initial guess for a Gaussian 2-D fitting (using CASA {\em imfit} algorithm) in order to derived accurate fluxes from the
extracted sources\footnote{We found that {\em getsources} tend to underestimate the recovered fluxes from our data}.
Also, we kept the same source nomenclature used by  \citet{Sridharan2014}, but adding numbers where higher multiplicity
was discovered with respect to the SMA map.
Using the standard procedure to calculate masses
from dust emission \citep{Hildebrand1983} we computed masses for all 15 sources in our catalog
assuming a dust opacity of $\kappa_{\mathrm{1.3 mm}} = 0.01$ cm$^{2}$g$^{-1}$ \citep{Ossenkopf1994},
a gas to dust ratio of 1:100, and  
a dust temperature of $T_{\mathrm{dust}} = 25$ K \citep{Bally2010}, with the exception of the
hot core, source A, where we used a range between $70 < T_{\mathrm{dust}} < 150$  K. Although the SMA detected CH$_{3}$CN emission 
towards B1 and C, it was unresolved and not sufficient to derived temperatures; hence, we used $T_{\mathrm{dust}} = 25$ K
for these sources. \citet{Herpin2012} modeled an SED and derived a temperature profile for source A using all the publicly 
available data on W43-MM1 to date.
Their model suggests a temperature of $\sim$ 150 K at 2500 AU distance (0.5$^{\prime \prime}$ radial; 1$^{\prime \prime}$ size)
and $\sim 70$ K at about 8000 AU distance (0.7$^{\prime \prime}$ radial; 1.4$^{\prime \prime}$ size), which
are the length-scales sampled by ALMA.
However if larger spatial scales than our source size are considered, the temperature might be lower and in
the order of 30 K \citep[as suggested by][]{Bally2010}. These scales ($>10^{\prime \prime}$) are consistent with
Herschel primary beam at 160 $\mu$m and, off course, larger than our deconvolved source sizes and synthesized beam.
Therefore and using this temperature range, our derived mass for source A is between 312 and  146~ \Msun 
(with the exception of the 30 K temperature derived mass of 728~\Msun\ ).
These estimates put source A below other massive clumps such as
the SDC335-MM1 mass estimate of 545~ \Msun\ \citep{Peretto2013} and G31.41+0.31 with a mass estimate of 577~\Msun\
 \citep{Girart2009}\footnote{Note that in this massive core a magnetic field
strength of 10 mG was derived using n(H$_{2}$)=$3\times 10^{6}$ cm$^{-3}$}. However, these  mass estimates where derived from observations
sampling larger length scales than our ALMA W43 observations. The deconvolved size of SDC335-MM1 is about 0.054
which is 2 times the size of source A or 4 times the area. A simple estimate assuming a $r^{-2}$ density profile,
will give about 272~\Msun\ per source A size for SDC335-MM1 and about 205~\Msun\ per source A size for G31.41+0.31. 

\section{THE MAGNETIC FIELD MORPHOLOGY}\label{se:MF}

\begin{figure}
\centering
\includegraphics[width=0.95\hsize]{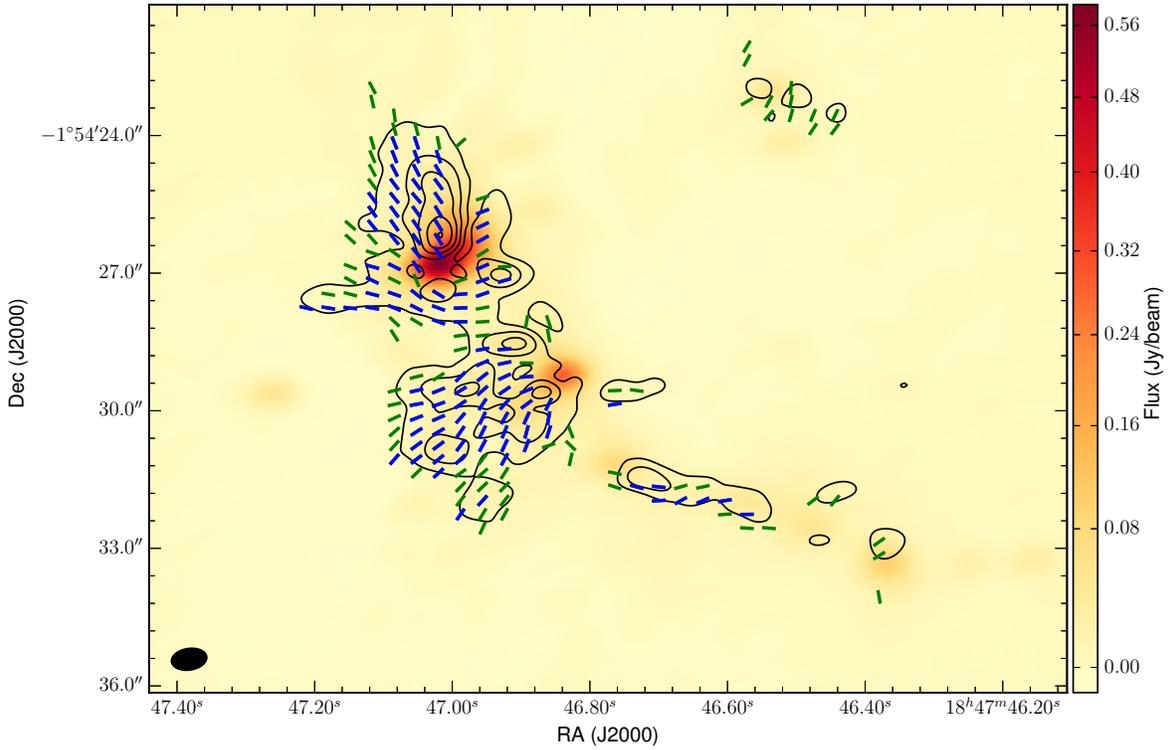}\caption{The Figure shows the magnetic field morphology over W43-MM1.
    The Stokes I emission is shown as color scale, the dust, rician debiased, polarized intensity is shown in contours of
    0.55,  1.3,  2.0,  2.7, and 3.5 mJy $\mathrm{beam}^{-1}$. The magnetic field morphology is shows as 
    pseudo-vectors at a significance of 3$\sigma$ in green and 5$\sigma$ in blue, where $\sigma = 93 \mu$Jy $\mathrm{beam}^{-1}$ 
    corresponds to the noise in the polarized intensity image. The length of each
    pseudo-vectors is normalized. Also, each pseudo-vector is plotted every half-beam, i.e. in steps of 8 and 4 pixels, where
    the beam is $13\times7$ pixels.
  }
\label{w43_pol}
\end{figure}

\noindent The polarized emission from W43-MM1 shows fractional polarization levels between 0.03\% to 22\% where the lowest values
are seen over the peaks in Stokes I, which corresponds to the well known polarization-hole \citep[and references therein]{Hull2014}. 
Assuming perfect grain alignment, the magnetic 
field morphology onto the plane of the sky is inferred from the polarized emission by rotating the electric vector position angle (EVPA) by 90$^{\circ}$
 and shown in Figure \ref{w43_pol}.
The field morphology shows a smooth and ordered pattern over the filament on angular scales $< 15^{\prime\prime}$
\footnote{The ALMA polarization accuracy of 0.1\% in fractional polarization is only guaranteed
within 1/3 of FWHM, which in our data this corresponds to $\sim 10^{\prime \prime}$. However, recent 
results from ALMA commissioning \citep{Cortes2015} showed that the systematic error within -3 dB level 
($\sim 15^{\prime \prime}$) of the primary beam is less than 0.5\% in band 6,
which is larger than the size of the W43-MM1 filament. In fact, our observations are  consistent to the SMA
data beyond the 1/3 of FWHM or 10$^{\prime \prime}$ limit.},
where  ordered magnetic fields are found in massive dense cores when 
observed at similar spatial scales, as it has been  shown from a relatively large sample by \citet{Zhang2014b}.
We found overall agreement with the SMA results, but now given our higher resolution and sensitivity, we can
see the field morphology in greater detail. In fact 
for source A, there is a $\sim$90$^{\circ}$ change in orientation to the west of the clump, where there is also
a decrease in polarized intensity, and  towards D1, the field
changes about $90^{\circ}$ in orientation with respect to the main filament.
For analysis, we divided the filament into four regions according to the boundaries in the field pattern.
Although, the field morphology is continuous from A to B1, we 
set the boundary to the north of B2 where we plotted the 
lowest contour in the polarized intensity. We did that in order to analyze the field locally to source A and sources B1,
B2, B3, B4, and E.

\noindent To understand the dynamical importance of the magnetic field over W43-MM1, we estimated the strength of
the field using the Chandrasekhar \& Fermi technique 
\citep[here after CF][]{Chandrasekhar1953} as follows from \citet{Crutcher2004}. 

\begin{equation}
    \label{cf}
    B_{\mathrm{pos}} = 9.3 \frac{\sqrt{n_{\mathrm{H_{2}}}}\Delta V}{\delta \phi}
\end{equation}

\noindent where $n_{\mathrm{H_{2}}}$ is the molecular Hydrogen number density calculated
as an average  over the region sampled by the polarized dust emission, $\Delta V$ is the FWHM
from a line tracing the gas motions in W43 which in this case was taken to be 3.0 km/s 
from H$^{13}$CO$^{+}$(4-3)
and DCO$^{+}$(5-4) single dish emission \citep{Cortes2010,Cortes2011}, and $\delta \phi$ is the EVPA dispersion,
which we calculated as the standard deviation of the EVPAs for each region. 
Thus, the derived field strengths are local to all the sources in a particular region. 
Table \ref{tab2} summarizes our polarization results with the corresponding estimations for the field. 

\noindent The usage of the CF method have been debated given the small angle dispersion approximation required and the
assumption of energy equipartition. From Table \ref{tab2} we see that 2 out of 4 of our regions have EVPA dispersions
larger than the $25^{\circ}$ limit suggested by \citet{Ostriker2001}.
Given this, we included modified versions of the CF method \citep{Heitsch2001,Falceta2008},
to calculate the field strength on each region independently. \citet{Heitsch2001} attempt to address the 
limitation of the small angle approximation by replacing $\delta\phi$ by $\delta \tan(\phi)$ which is
calculated locally and by adding a geometric correction to avoid underestimating the field in the super-Alfvenic case. 
In contrast, \citet{Falceta2008} assumed that the field perturbation is a global property and thus, they
replaced $\delta \phi$ by $\tan(\delta \phi) ~\sim 􏰅\delta B/B_{sky}$ in the denominator of equation \ref{cf}.
\noindent By using the 3 versions of the CF method, we obtained estimations of the magnetic field between 
0.2 to 7 mG for our 4 regions, where for source A we obtained values between 1 and 4 mG. \citet{Crutcher2012} plotted the
most up-to-date field profile from Zeeman measurements. 
Figures 6 and 7 in that work, suggests that our estimates for B$_{\mathrm{pos}}$ are
consistent with the curve if we extrapolate the profile as, unfortunately, the plot lacks field values for the $10^{25}$ cm$^{-2}$ 
range in column density.

\section{SUMMARY AND DISCUSSION}\label{se:sum}

\noindent The ALMA results on W43\_MM1 suggest that we are seeing a highly fragmented filament where the emission
is dominated by a single clump, source A. 
Indeed, from Table \ref{tab1} 
we see that the bulk of fragments (13) have fluxes between 32 and 200 mJy; while source A alone is about 2 Jy.
The high degree of fragmentation seen in the ALMA data pose the question about the gravitational
stability of these sources.
Although we do not have high resolution line data to address the kinematics
and determine if these sources are self-gravitating, the infalling motions detected by 
\citet{Cortes2010} (covering an area 
of $48^{\prime \prime} \times 48^{\prime \prime}$) are likely
showing accretion from larger scales onto the whole filament and suggesting gravitational collapse.
To test the gravitational stability of each 
fragment, we calculated the thermal Jeans
length as $\lambda_{J} = c_{s}(\pi/G\rho)^{1/2}$, where $c_{s}=\sqrt{KT/m_{\mathrm{H_{2}}}}$ the sound speed, $\rho$ is the
volume density, and $G$ is the gravitational constant.
We found that our deconvolved sizes are larger than the estimated Jeans lengths by roughly
an order of magnitude for most fragments. Thus, we cannot conclude what fraction of these fragments
are gravitationally unstable from these data alone. Interestingly is the case of source A which
has been determined to be gravitationally bound \citep[and references therein]{Louvet2014,Cortes2010}
but its deconvolved size is between a factor of 10 and 6 larger than its Jeans length. If the Jeans length
suggests further fragmentation, it is possible that source A has a larger multiplicity of fragments which 
requires higher angular resolution to resolve.

\noindent Using the derived magnetic field estimations,
we can compute the mass to magnetic flux ratio for all our sources using our B$_{\mathrm{pos}}$ estimations.
We do this following \citet{Crutcher2004} as,

\begin{equation}
\lambda_{B} = 7.6\times10^{-21}\frac{N(\mathrm{H_{2}})}{3\mathrm{B}_{\mathrm{pos}}},
\end{equation}

\noindent where N$(\mathrm{H_{2}})$ is molecular hydrogen column density in cm$^{-2}$
calculated for each source independently, 
B$_{\mathrm{pos}}$ is the magnetic field strength in $\mu G$ assumed to be unique for a given region, 
and factor of 3 in the denominator corresponds to 
an statistical geometrical correction \citep{Crutcher2004}. 
This statistical geometrical factor of 3 is for a uniform density slab and will be smaller than 3 for more physically 
realistic cases, such as a centrally condensed disk.
We found highly super-critical values for all fragments in the filament (see Table \ref{tab2}), 
showing that the field is not strong  enough to support them against gravity. 
Also, it is worth noting that polarized emission decreases significantly towards the dust peaks in W43-MM1,
which has already been observed towards DR21(OH) by the SMA \citep{Girart2013}.
This is particularly evident towards source A where the polarized intensity emission goes
below the 5$\sigma$ level close the dust peak.
Additionally, the B$_{\mathrm{pos}}$ morphology in source A seems to indicate the field weakness as the
lines appear to be bent towards the dust peak.
This additional evidence, suggests that gravity is pulling the field lines as the gas is infalling
due to gravitational collapse.
However, what fraction of these fragments are bound by gravity is 
still open; and thus, further fragmentation cannot be ruled out.

\noindent Here, we have presented ALMA polarized dust emission observations of W43-MM1.
We have found a fragmented filament threaded by a smooth, ordered, and highly detailed magnetic 
field. We have derived mass estimates for 15 fragments extracted from the continuum map, showing that 
source A dominates the flux distribution with a 51\% of the total flux. Using temperature modeling by 
others, we derived a mass range for source A between 146 and 312 ~\Msun\ under the length-scales sampled by ALMA.
However, it is not clear if 
further fragmentation is ongoing inside source A and only higher resolution ALMA observations can reveal this.
We have derived magnetic field strengths for all fragments with sufficient polarization data and found
field estimations in range of 0.2 - 7 mG using 3 versions of the CF method. 
Derivation of the mass to magnetic flux ratio
indicate that the fragments are super-critical suggesting that the field is dominated by gravity
at this stage of the W43-MM1 evolution.

\begin{acknowledgements}
\noindent The National Radio Astronomy Observatory is a facility of the National Science Foundation operated under cooperative agreement by Associated Universities, Inc.\\
P.C.C. would like to thank Ed Fomalont for very helpful and enlighten discussion about polarization data analysis.\\
J.M.G. acknowledges support from MICINN AYA2014-57369-C3-P and the MECD PRX15/00435 grants (Spain).\\
Z.Y.L. is supported in part by NASA NNX14AB38G and NSF AST-1313083.\\
S.P.L. thanks the support of the Ministry of Science and Technology (MoST) of Taiwan through
Grants NSC 98-2112-M-007-007-MY3, NSC 101-2119-M-007-004, and MoST 102-2119-M-007-004- MY3.\\
This research made use of APLpy, an open-source plotting package for Python hosted at http://aplpy.github.com
\end{acknowledgements}

\bibliographystyle{apj}


\begin{deluxetable}{c c c c c c c c c c c c}        
\tablecolumns{12}
\tablewidth{0pt}
\tabletypesize{\scriptsize}
\tablecaption{Continuum Sources\label{tab1}}        
\tablehead{
\colhead{ID} & 
\colhead{R.A.} &
\colhead{DEC} & 
\colhead{Peak\tablenotemark{a}} & 
\colhead{Flux} & 
\colhead{Major\tablenotemark{b}} & 
\colhead{Minor\tablenotemark{b}} & 
\colhead{P.A.} & 
\colhead{n\tablenotemark{c,d}} & 
\colhead{Mass\tablenotemark{c}} & 
\colhead{Size\tablenotemark{e}} & 
\colhead{$\lambda_{J}$\tablenotemark{c}}\\
\colhead{ } & 
\colhead{[J2000]} & 
\colhead{[J2000]} & 
\colhead{[mJy beam$^{-1}$]} & 
\colhead{[mJy]} & 
\colhead{[arcsec]} & 
\colhead{[arcsec]} & 
\colhead{[deg]} & 
\colhead{[$10^{7}$ cm$^{-3}$ ]} & 
\colhead{[\Msun]} & 
\colhead{[mpc]} & 
\colhead{[mpc]} 
}
\startdata
A & 18:47:47.0 &  -01:54:26.7 & 503.1 $\pm$  26 & 1988.2 & 1.30 & 0.80 & 131.4 &  38,81.3,190 & 146,312,729 & 27.09 & 5,3.4,2.2\\
B1 & 18:47:46.8 &  -01:54:29.3 & 271.4 $\pm$  15 & 504.1 & 0.58 & 0.51 & 138 & 384.0 & 222 & 14.4 & 1.5\\
C & 18:47:46.4 &  -01:54:33.4 & 95.2 $\pm$   3 & 190.5 & 0.67 & 0.46 &  33 & 133.9 &  84 & 14.8 & 2.6\\
H & 18:47:46.8 &  -01:54:31.2 & 73.5 $\pm$   4 & 151.9 & 0.72 & 0.50 &  60 & 85.0 &  67 & 16.0 & 3.3\\
B4 & 18:47:46.9 & -01:54:30.1 & 49.8 $\pm$   8 & 121.1 & 1.00 & 0.60 &  72 & 31.3 &  53 & 20.7 & 5.4\\
G & 18:47:47.3 &  -01:54:29.6 & 48.1 $\pm$   2 & 48.4 & 0.77 & 0.47 & -81 & 26.6 &  21 & 16.1 & 5.9\\
B2 & 18:47:46.9 &  -01:54:28.6 & 46.1 $\pm$   5 & 201.3 & 1.48 & 0.82 & 103 & 17.9 &  89 & 29.5 & 7.1\\
F1 & 18:47:46.5 &  -01:54:23.1 & 44.7 $\pm$   1 & 71.9 & 0.77 & 0.47 & -81 & 39.4 &  32 & 16.1 & 4.8\\
B3 & 18:47:47.0 &  -01:54:29.6 & 43.7 $\pm$   4 & 133.3 & 1.10 & 0.68 & 101 & 24.9 &  59 & 23.0 & 6.1\\
D2 & 18:47:46.5 &  -01:54:32.4 & 42.6 $\pm$   4 & 153.9 & 1.26 & 0.74 &  75 & 20.8 &  68 & 25.7 & 6.6\\
E & 18:47:47.0 &  -01:54:30.8 & 38.6 $\pm$   1 & 84.0 & 0.73 & 0.51 & 158 & 43.8 &  37 & 16.3 & 4.6\\
D1 & 18:47:46.6 &  -01:54:32.0 & 36.7 $\pm$   2 & 137.1 & 1.61 & 0.58 &  79 & 18.5 &  60 & 25.7 & 7.0\\
K & 18:47:46.2 &  -01:54:33.3 & 25.3 $\pm$   1 & 40.4 & 0.77 & 0.47 & -81 & 22.2 &  18 & 16.1 & 6.4\\
F2 & 18:47:46.5 &  -01:54:24.1 & 24.3 $\pm$   2 & 32.4 & 0.77 & 0.47 & -81 & 17.8 &  14 & 16.1 & 7.2\\
J & 18:47:46.3 &  -01:54:33.4 & 15.6 $\pm$   1 & 37.2 & 0.93 & 0.54 & 111 & 12.7 &  16 & 18.8 & 8.5\\

\enddata
\tablenotetext{a}{Sources are sorted according to their peak flux}
\tablenotetext{b}{Deconvolved major and minor axes are obtained from the Gaussian fitting procedure.}
\tablenotetext{c}{For the hot core (source A), we are showing calculations for a $T_{\mathrm{dust}}$=30, 70, and 150  K}
\tablenotetext{d}{Volume number densities were calculated assuming spherical geometry.}
\tablenotetext{e}{The fragment size is calculated as d = 5500*$\sqrt{b_{maj} \times b_{min}}$ [pc].}
\end{deluxetable}

\begin{deluxetable}{c c c c c c c c c c c c c c c}        
\tablecolumns{15}
\tablewidth{0pt}
\tabletypesize{\scriptsize}
\tablecaption{Polarization data and magnetic field estimations onto the plane of the sky. All parameters are dervied assuming a temperature of 25 K with the exception of source A, where we are showing values for $T_{\mathrm{dust}} = 70$ and 150 K. The polarization statistics are calculated from the 5$\sigma$ data\label{tab2}}        
\tablehead{
    \colhead{Source} & 
    \colhead{Region} & 
    \colhead{n$_{\mathrm{r}}$\tablenotemark{a}} &
    \colhead{N\tablenotemark{b}}& 
    \colhead{$<\phi>$\tablenotemark{c}} & 
    \colhead{$\delta \phi$\tablenotemark{c}} & 
    \colhead{F$_{min}$\tablenotemark{c}} & 
    \colhead{F$_{max}$\tablenotemark{c}} & 
    \colhead{$<\mathrm{F}>$\tablenotemark{c}} & 
    \colhead{B$_{1}$\tablenotemark{d}} & 
    \colhead{B$_{2}$\tablenotemark{e}} & 
    \colhead{B$_{3}$\tablenotemark{f}} & 
    \colhead{$\lambda_{\mathrm{B_{1}}}$\tablenotemark{g,b}} & 
    \colhead{$\lambda_{\mathrm{B_{2}}}$\tablenotemark{h,b}} &
    \colhead{$\lambda_{\mathrm{B_{3}}}$\tablenotemark{i,b}}   \\ 
    \colhead{ } & 
    \colhead{ } & 
    \colhead{[$10^{7}$ cm$^{-3}$ ]} & 
    \colhead{[$10^{24}$ cm$^{-2}$ ]} & 
    \colhead{[$^{\circ}$]} & 
    \colhead{[$^{\circ}$]} & 
    \colhead{[\%]} & 
    \colhead{[\%]} & 
    \colhead{[\%]} & 
    \colhead{[mG]} & 
    \colhead{[mG]} & 
    \colhead{[mG]} &
    \colhead{ } & 
    \colhead{ } & 
    \colhead{ } 
}
\startdata 
A & 1 &  1.4 & 68.0-31.7 & -30.5 & 36.2 & 0.41 & 13.9 & 4.9 &   3 &   4 &   1 & 59-27 & 40-19 & 145-68\\
B1 & 2 &  1.5 & 170.7 & 42.5 & 21.9 & 0.46 & 19.6 & 6.2 &   5 &   8 &   2 &  89 &  55 & 225 \\
E & 2 &  1.5 & 22.1 & 42.5 & 21.9 & 0.46 & 19.6 & 6.2 &   5 &   8 &   2 &  12 &   7 &  29 \\
B2 & 2 & 1.5 & 16.3 & 42.5 & 21.9 & 0.46 & 19.6 & 6.2 &   5 &   8 &   2 &   9 &   5 &  21 \\
B3 & 2 & 1.5 & 17.7 & 42.5 & 21.9 & 0.46 & 19.6 & 6.2 &   5 &   8 &   2 &   9 &   6 &  23 \\
B4 & 2 & 1.5 & 20.0 & 42.5 & 21.9 & 0.46 & 19.6 & 6.2 &   5 &   8 &   2 &  10 &   6 &  26 \\
C & 3 & 1.1 & 61.2 & 12.8 & 49.7 & 0.87 & 9.4 & 4.1 &   2 &   2 &   0.2 &  84 &  67 & 701 \\
H & 3 &  1.1 & 41.9 & 12.8 & 49.7 & 0.87 & 9.4 & 4.1 &   2 &   2 &   0.2 &  57 &  46 & 480 \\
D2 & 3 &  1.1 & 16.4 & 12.8 & 49.7 & 0.87 & 9.4 & 4.1 &   2 &   2 &   0.2 &  22 &  18 & 189 \\
D1 & 3 &  1.1 & 14.7 & 12.8 & 49.7 & 0.87 & 9.4 & 4.1 &   2 &   2 &   0.2 &  20 &  16 & 168 \\
F1 & 4 &  0.4 & 19.6 & 54.3 & 22.0 & 2.30 & 22.4 & 6.8 &   3 &   4 &   1 &  19 &  12 &  53 \\
F2 & 4 &  0.4 & 8.8 & 54.3 & 22.0 & 2.30 & 22.4 & 6.8 &   3 &   4 &   1 &   8 &   5 &  24 \\
\enddata

\tablenotetext{a}{The number density used to estimate B$_{\mathrm{pos}}$, calculated from the Stokes I emission across the entire region}
\tablenotetext{b}{For the hot core, the estimations are calculated using a temperature range of  70 and 150 K}
\tablenotetext{c}{Here $<\phi>$ is the average EVPA, $\delta \phi$ is the EVPA dispersion (calculated using circular statistics), F$_{min}$ is the minumum fractional polarization,
F$_{max}$ is the maximun fractional polarization, and $<\mathrm{F}>$ is average fractional polarization value. All values are computed for the region indicated in column 2.}
\tablenotetext{d}{Estimations of the magnetic field, in the plane of the sky, done with the original CF method (see equation \ref{cf} in the text)}
\tablenotetext{e}{Estimations of the magnetic field, in the plane of the sky, done using the corrections implemented by \citet{Falceta2008} equation 9}
\tablenotetext{f}{Estimations of the magnetic fieldin the plane of the sky,  done using the corrections implemented by \citet{Heitsch2001} equation 12}
\tablenotetext{g}{Mass to magnetic flux estimate using field strength estimate B$_{1}$}
\tablenotetext{h}{Mass to magnetic flux estimate using field strength estimate B$_{2}$}
\tablenotetext{i}{Mass to magnetic flux estimate using field strength estimate B$_{3}$}
\end{deluxetable}
\end{document}